\documentclass[journal=inoraj,manuscript=article]{achemso}
\usepackage{graphicx,bm}
\usepackage{dcolumn}
\usepackage{bm}
\usepackage{upgreek}
\usepackage{times}
\usepackage[caption=false]{subfig}
\usepackage{mathrsfs}
\usepackage{amsmath} 
\usepackage[version=3]{mhchem} 


\title{Magnetic and structural studies of G-phase compound Ni$_{16}$Mn$_6$Si$_7$}

\author{S. J. Ahmed}
\affiliation{Department of Materials Science and Engineering, McMaster University, Hamilton, Ontario.}

\author{J. E. Greedan}
\affiliation{Department of Chemistry, McMaster University, Hamilton, Ontario.}
\alsoaffiliation{Brockhouse Institute of Materials Research, McMaster University, Hamilton, Ontario.}

\author{C. Boyer}
\affiliation{Canadian Neutron Beam Centre, Canadian Nuclear Laboratories, Chalk River, Ontario.}
\author{M. Niewczas}
\affiliation{Department of Materials Science and Engineering, McMaster University, Hamilton, Ontario.}
\alsoaffiliation{Brockhouse Institute of Materials Research, McMaster University, Hamilton, Ontario.}
\email{niewczas@mcmaster.ca}

\date{\today}

\begin{document}
\begin{abstract}

Transition metal compounds with complex crystal structures tend to demonstrate interesting magnetic coupling resulting in unusual magnetic properties.  In this work, the structural and magnetic characterization of a single crystal of the Ni-Mn-Si based G-phase compound, Mn$_6$Ni$_{16}$Si$_7$, grown by the Czochralski method, is reported. In this structure isolated octahedral Mn$_6$ clusters form a f.c.c. lattice. As each octahedron consists of eight edge-sharing equilateral triangles, the possibility for geometric frustration exists. Magnetization and specific heat measurements showed two magnetic phase transitions at 197 K and 50 K, respectively. At 100 K neutron diffraction on powder samples shows a magnetic structure with k = (001) in which only four of the six Mn spins per cluster order along $<100>$ directions giving a two dimensional magnetic structure consistent with intra-cluster frustration. Below the 50 K phase transition the Mn spins cant away from $<100>$ directions and a weak moment develops on the two remaining Mn octahedral sites.

 \end{abstract}

\maketitle

\section{Introduction}\label{sec:Introduction}

The geometry of the crystal structure and strong nearest neighbour interactions in compounds containing transition metal ions give rise to phenomena such as geometric frustration, noncollinear magnetism and the coupling of these properties \cite{Fries1997,Hur2004,Ye2007,Melchy2009,Deng2015}. The compounds display new magnetic properties including canted ferromagnetic state \cite{Fruchart1978,Ramirez1994,Greedan2001,Yu2003,Greedan2006,Yang2011,Matan2011}, non collinear ferrimagnetism \cite{Yan2014}, metamagnetism \cite{Dendrinou-Samara2018}, giant magnetostriction \cite{Shimizu2012,Shibayama2011}, unusual magnetic hysteresis behaviour \cite{Sun2014}, giant magnetoresistance \cite{Kamishima2000} among others. The enhanced magnetic interactions make these materials prominent candidates for applications in quantum computation, data storage, magnetic refrigeration, and spintronics~\cite{Stamp2009,Evangelisti2010,Troiani2011,Timco2011,Dong2017}.\\

The G-Phase based compounds crystalize in a Mg$_6$Cu$_{16}$Si$_7$ (ternary) or Th$_6$Mn$_{23}$ (binary) type structure. The system consists of a complex network of octahedra and supertetrahedra with at least one of the constituting element being a transition metal~\cite{Hellenbrandt2004,Holman2008}. Several studies of these systems have focused primarily on the properties associated with high density magnetic information recording media, magnetic suppression and superconductivity~\cite{Spiegel1963,Chaudouet1983,Bowden1997,Ostorero2001,Ostorero2005,Holman2008}. However, a complex crystal structure containing transition metals holds promises for unique magnetic coupling that can give rise to multifunctional behaviour in these compounds.\\

The Ni-Mn-Si based Mn$_6$Ni$_{16}$Si$_7$ type G phase was proposed to be an antiferromagnetic system below 200 K by~\citet{STADNIK1983}. Further neutron diffraction and magnetization analysis by \citet{Kolenda1991} resulted in a proposed magnetic structure in which all Mn spins are ordered below 200 K with a moment of $2.7 \mu_B$/Mn  atom at 80 K, the base temperature of their study. Nonetheless, some aspects of the physical properties of the system seem inconsistent with this model, for example the magnetic susceptibility increases as the temperature decreases below T$_{N}$ which deviates  from  conventional antiferromagnetic behavior in a polycrystalline system.\\

In the present work, comprehensive studies of the magnetic properties of a Mn$_6$Ni$_{16}$Si$_7$ single crystal grown by the Czochralski method as well as neutron diffraction on a powder sample are presented. The temperature range of the measurements is extended to 2 K. The emergence of an additional low-temperature phase transition below 50 K is reported. Magnetic structures refined at 100 K and 3.5 K from the neutron data are found to be more consistent with the bulk susceptibility data and the presence of geometric frustration. We discuss the possible changes in the magnetic structure of Mn$_6$Ni$_{16}$Si$_7$ compound in terms of the magnetic susceptibility measurements.

\section{Experimental details}\label{sec:Experimental details}

The Mn$_6$Ni$_{16}$Si$_7$ single crystal was grown by the Czochralski crystal growth method under an argon atmosphere using the RF heating. High purity Ni (99.95\%), Mn (99.98\%) and Si (99.999\%) were melted in an alumina crucible, and the crystal was pulled using a tungsten wire seed with a constant pulling rate of 0.5 mm/min and 30 rpm rotation. The growth direction was found to be along $<110>$ from Laue X-ray data. Pre-oriented samples of 2 mm$\times$2 mm$\times$2 mm were spark cut form the as-grown crystal and used for magnetic and specific heat measurements.\\

Studies of the magnetic properties were conducted with a Quantum Design MPMS SQUID magnetometer. The zero field cooled (ZFC)-field cooled (FC) magnetic susceptibly measurements were carried out from 350 K to 2 K on an oriented single crystal with a constant magnetic field of 1000 Oe applied in $<100>$, $<110>$ and $<111>$ direction.\\

The heat capacity of the compound was measured from 2K to 303 K using the Quantum Design PPMS system on the same 2 mm$\times$2 mm$\times$2 mm crystal samples used in magnetic studies.\\

The neutron diffraction studies of the Mn$_6$Ni$_{16}$Si$_7$ compound were performed at the Canadian Neutron Beam Centre in Chalk River on the C2 High Resolution Powder Diffractometer with a wavelength of 1.33{\AA}. The diffraction data were collected on about 4 grams of the powdered Mn$_6$Ni$_{16}$Si$_7$ sample, sealed in a thin-walled vanadium tube under argon atmosphere. The powder was obtained from a larger polycrystalline ingot, prepared by arc-melting of the pure elements under argon atmosphere. To improve homogeneity, the ingot was remelted three times. Excess Mn was added to compensate for the evaporation loss of the element. The compound was then sealed in an evacuated silica tube, annealed at 800 $^\circ$C for two weeks and subsequently, quenched in ice water mixture to further improve the crystallinity. No mass loss was observed due to annealing. Phase purity  was confirmed by X-ray powder diffraction using a \emph{PANalytical X'Pert Pro diffractometer} with Co K$\alpha_{1}$ radiation (Refinement profile shown in SI-1).\\

Refinement of the diffraction data was performed using the full-profile Rietveld method implemented in the FullProf program \cite{Rodriguez-Carvajal1990,Rodriguez-Carvajal2001,Rodriguez-Carvajal2011}. The magnetic configurations were generated with the representation analysis program SARAh \cite{Wills2000}.

\section{Results}\label{sec:Results}
\subsection{Magnetic susceptibility}\label{sec:Magnetic susceptibility}

\begin{figure}[htbp]
\subfloat[\label{sfig:ZFC-FC-1000-100}]{%
  \includegraphics[scale=0.32]{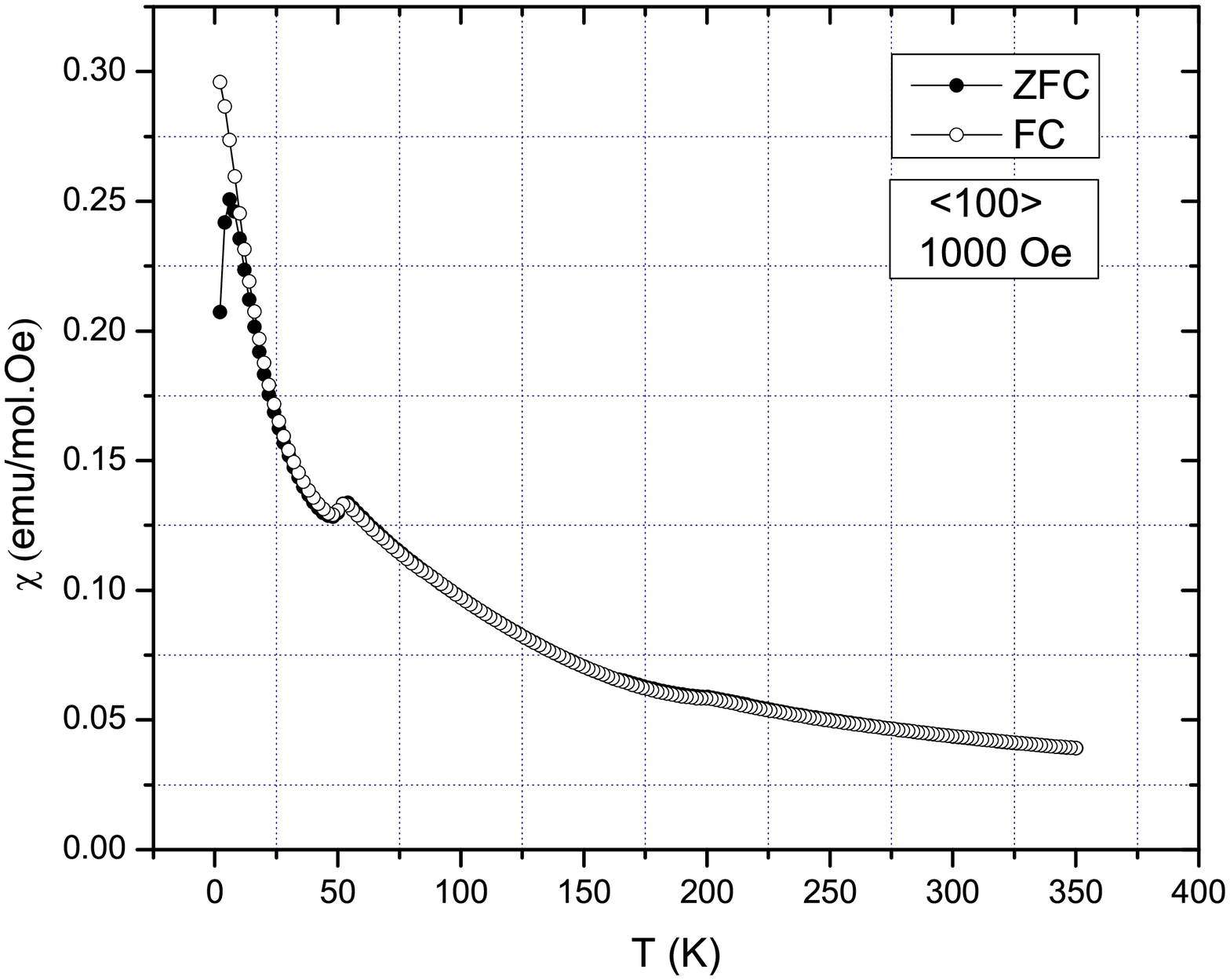}%
}
\subfloat[\label{sfig:ZFC-FC-1000-110C}]{%
  \includegraphics[scale=0.32]{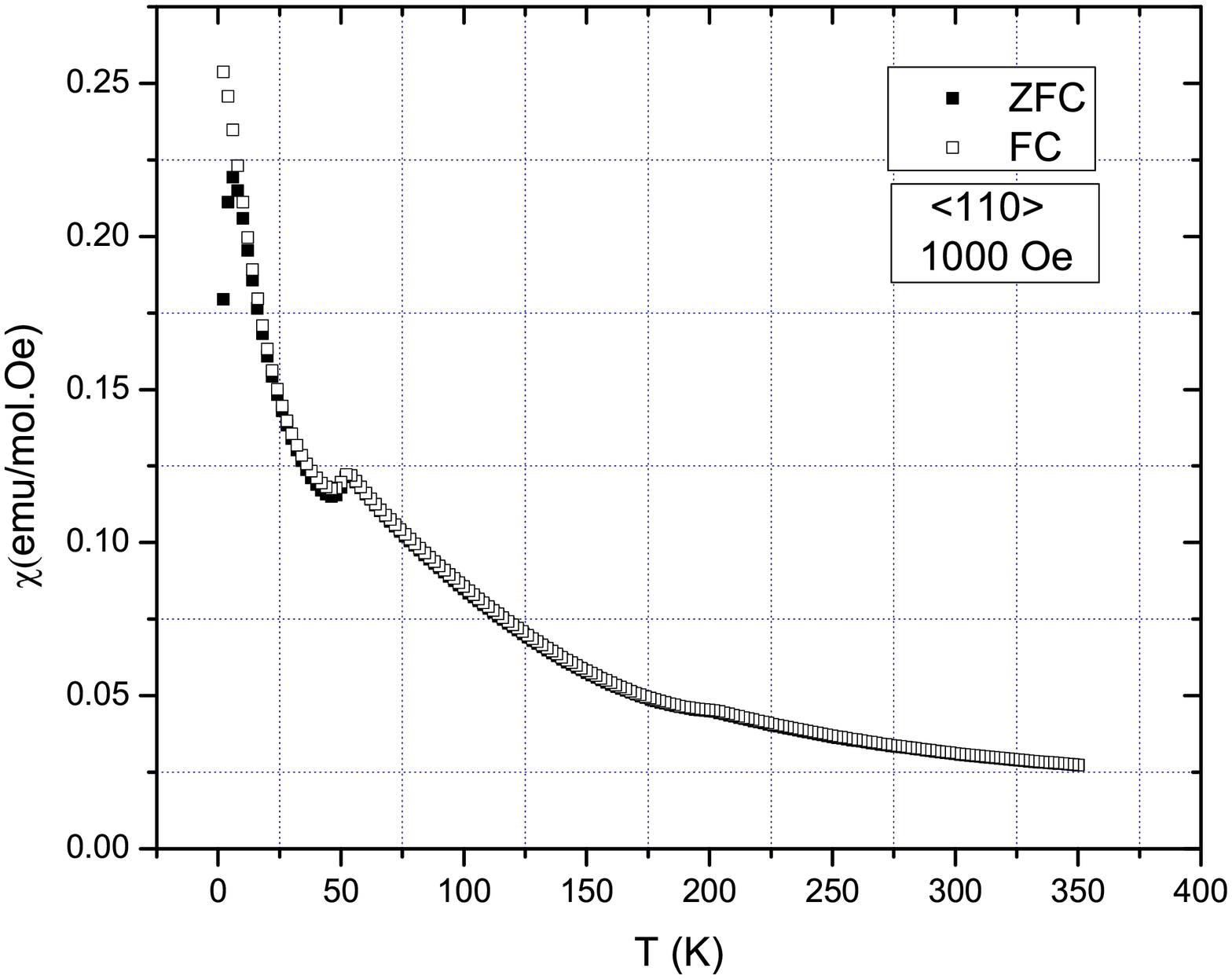}%
}\\
\subfloat[\label{sfig:ZFC-FC-1000-111}]{%
  \includegraphics[scale=0.32]{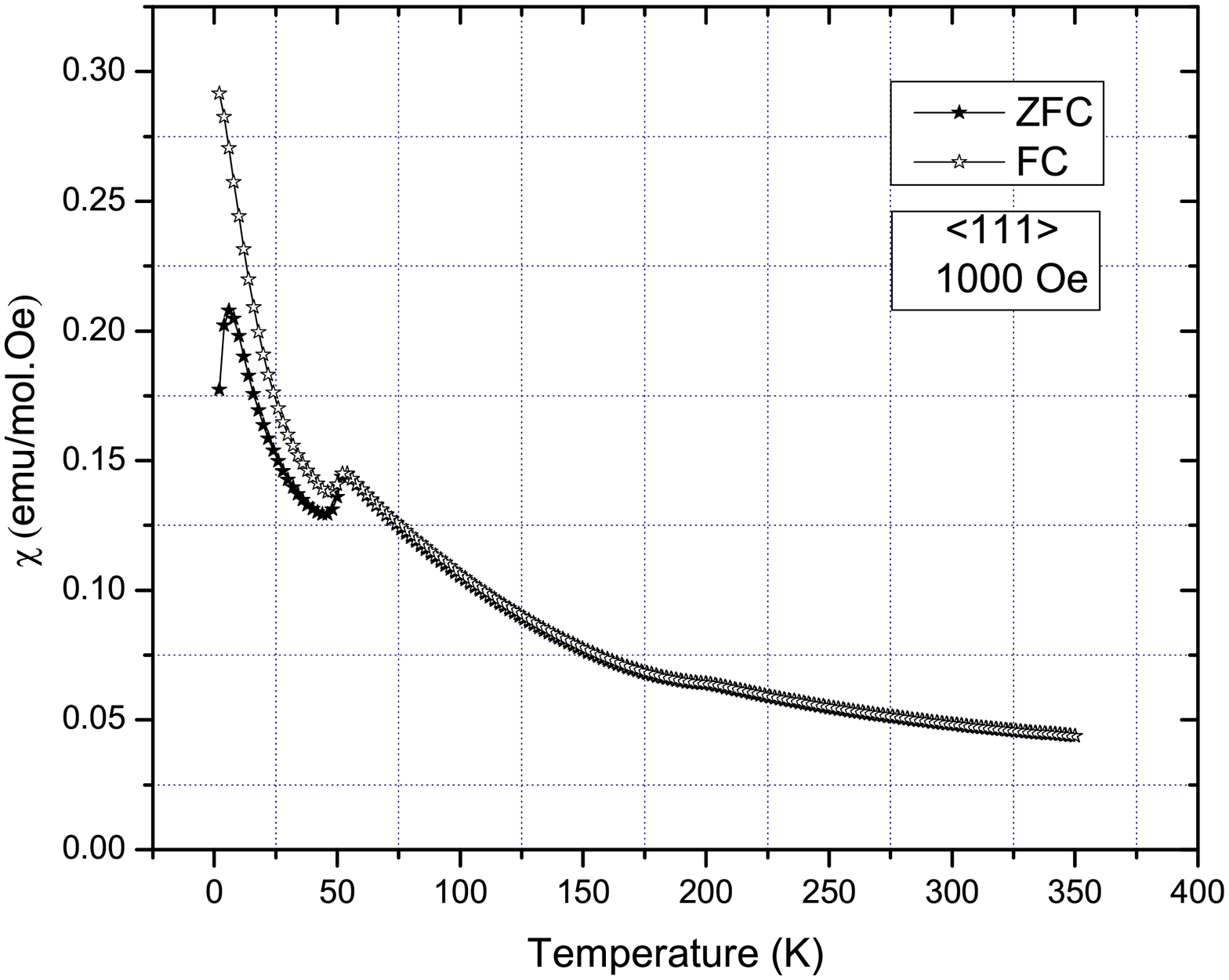}%
}
\caption{a) ZFC-FC magnetic susceptibility of Mn$_6$Ni$_{16}$Si$_7$ crystal at 1000 Oe field applied along a) $<100>$, b) $<110>$, and c) $<111>$.}
\label{fig:ZFC-FC-100-110-111-1000Oe}
\end{figure}

Fig.~\ref{fig:ZFC-FC-100-110-111-1000Oe} shows ZFC-FC magnetic susceptibly between 2 K and 350 K under the constant magnetic field of 1000 Oe applied along $<100>$, $<110>$ and $<111>$ directions of the crystal. The data clearly showed two transitions at $\sim$50 K and 197 K, the latter in agreement with \citet{Kolenda1991}. Additionally, there was a  ZFC/FC divergence below 6 K suggesting a weak spin freezing effect. Thus, two new anomalies are discovered here below the 80 K base temperature of the earlier studies. A comparison of the $<100>$, $<110>$ and $<111>$ ZFC-FC data (SI-2) showed a inconsistent anisotropic behavior of ZFC and FC magnetic susceptibility which indicates the applied 1000 Oe is too small to capture any magnetic anisotropy.

\begin{figure*}
\subfloat[\label{sfig:dchiTdt_100_1000Oe}]{%
  \includegraphics[scale=0.3]{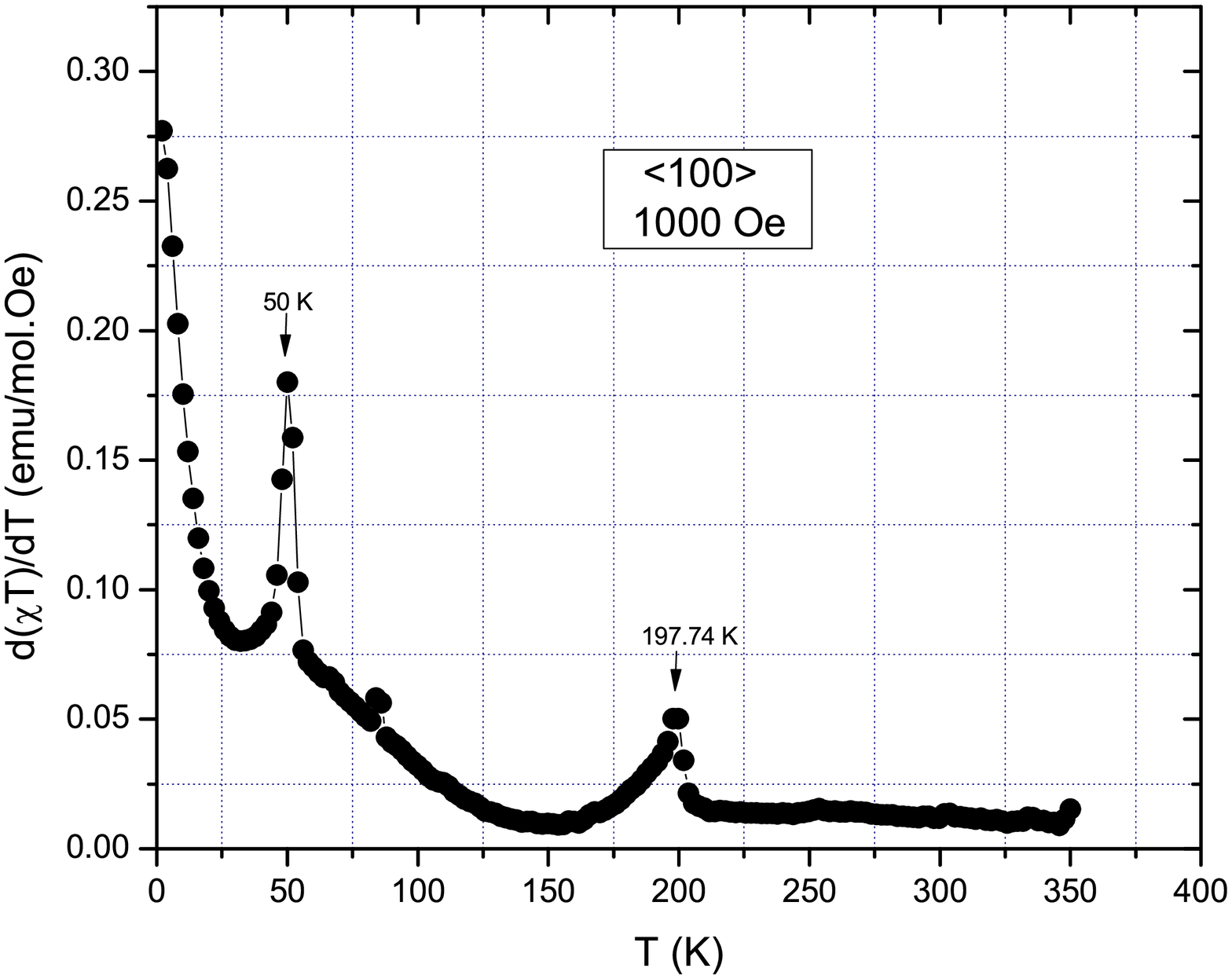}%
}
\subfloat[\label{sfig:FC_100_Curie_inverse}]{%
  \includegraphics[scale=0.3]{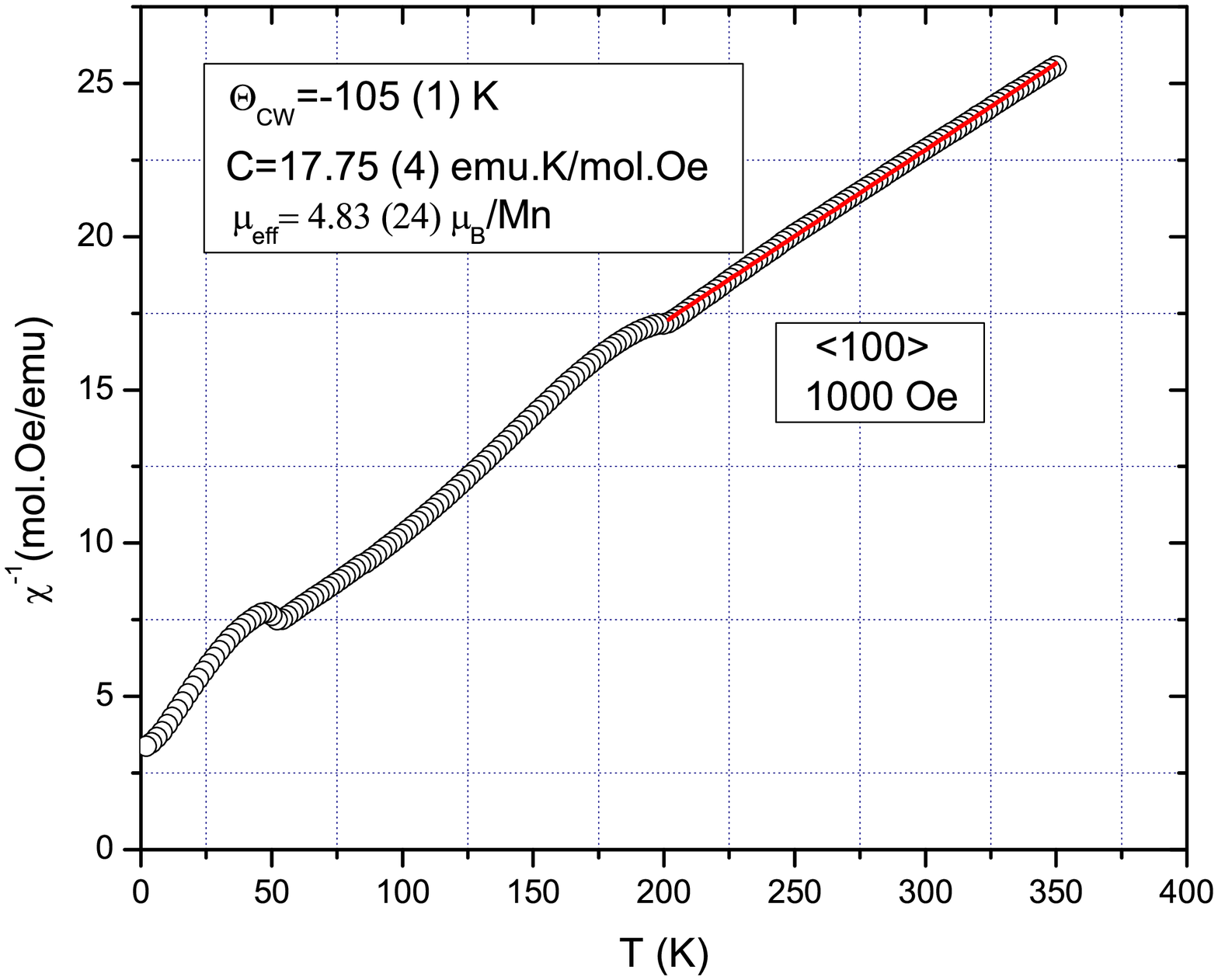}%
}
\caption{a)FC $d(\chi T)/dT$ vs. T plot identifying the transition temperatures, for crystal with the applied field along the $<100>$ direction. b) A Curie-Weiss fit on the inverse susceptibility data of Mn$_6$Ni$_{16}$Si$_7$ single crystals along the $<100>$ above 200 K.}
\label{Curie-Weiss}
\end{figure*}

Further analysis of the data with the field cooled Fisher heat capacity \cite{fisher1962relation}, $d(\chi T)/dT$ vs. T plot in Fig. \ref{sfig:dchiTdt_100_1000Oe}, showed two of the transitions (50 (1) K and 197 (1) K) quite clearly. The true N\'{e}el temperature was identified to be 197 (1) K from the second peak of the graph.

From the inverse FC susceptibly plot along the $<100>$ crystal direction, a Curie-Weiss fitting was done above 197 K (Fig.~\ref{sfig:FC_100_Curie_inverse}). The Curie constant, $C$ of 17.75 (4) emu.K/mol.Oe yielded an effective magnetic moment per Mn atom, $\mu_{eff}$/Mn of 4.83 (24) $\mu_B$ which is very close to the effective moment for of 4.90 $\mu_B$ for Mn$^{3+}$ (S=2) compared to the 5.91 $\mu_B$ for Mn$^{2+}$ (S=$\frac{5}{2})$. The Curie-Weiss parameters for FC susceptibility in other directions are listed in SI-table 1. Some caution should be taken as this fitting region may not be strictly in the paramagnetic regime. The Curie-Weiss temperature, $\Theta_{CW}$  were estimated to be -105 (1) K which indicates the presence of a strong antiferromagnetic exchange. Moreover, the susceptibility data continue to increase with decreasing temperature, indicating a persistent paramagnetic contribution which is likely due to the presence of geometrically frustrated paramagnetic ions.

\subsection{Heat Capacity measurement}\label{sec:Heat Capacity measurement}

\begin{figure}[htbp]
\subfloat[\label{sfig:Heat_Capacity_Comparison}]{%
  \includegraphics[scale=0.35]{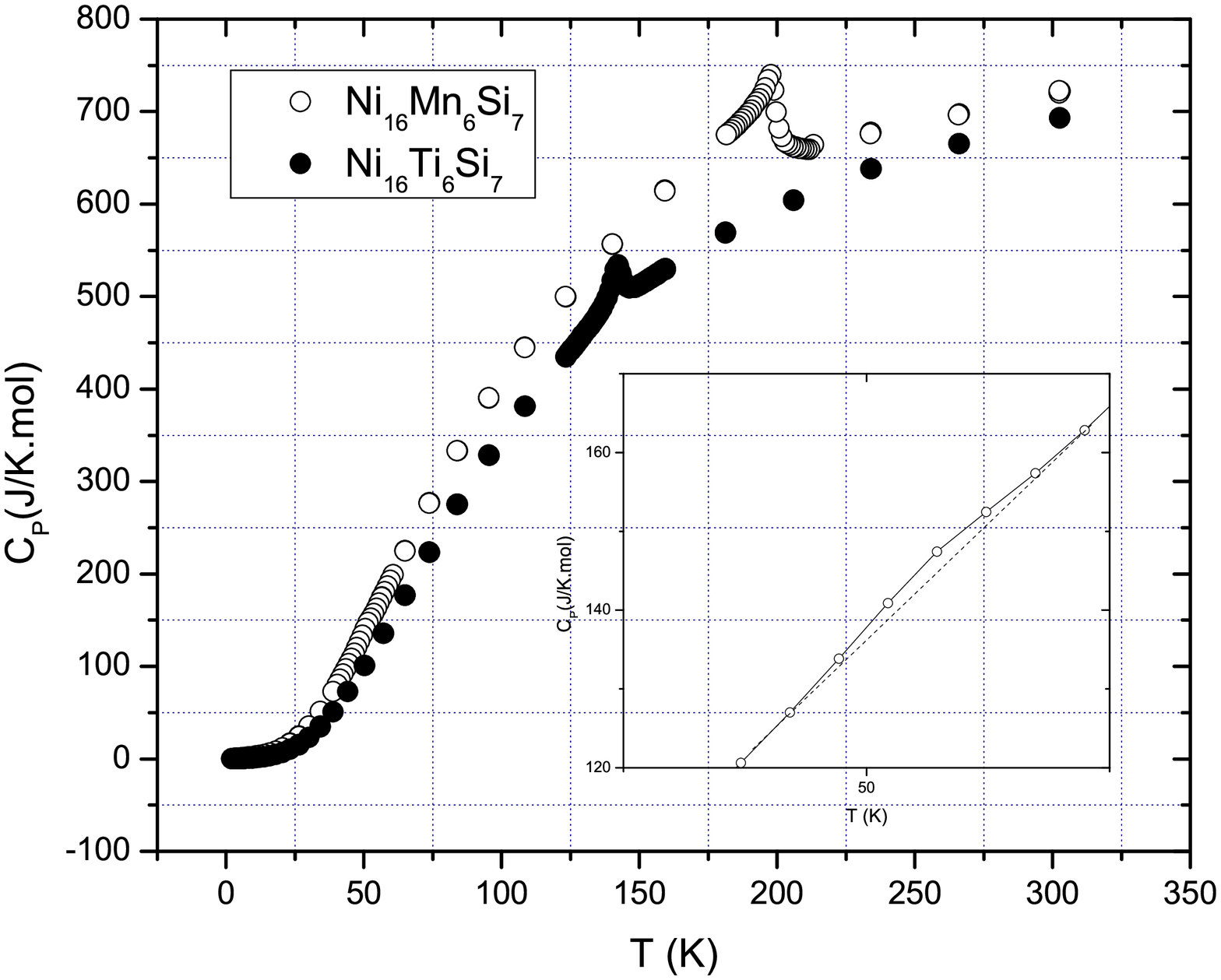}%
}\hfill
\subfloat[\label{sfig:CmagT}]{%
  \includegraphics[scale=0.35]{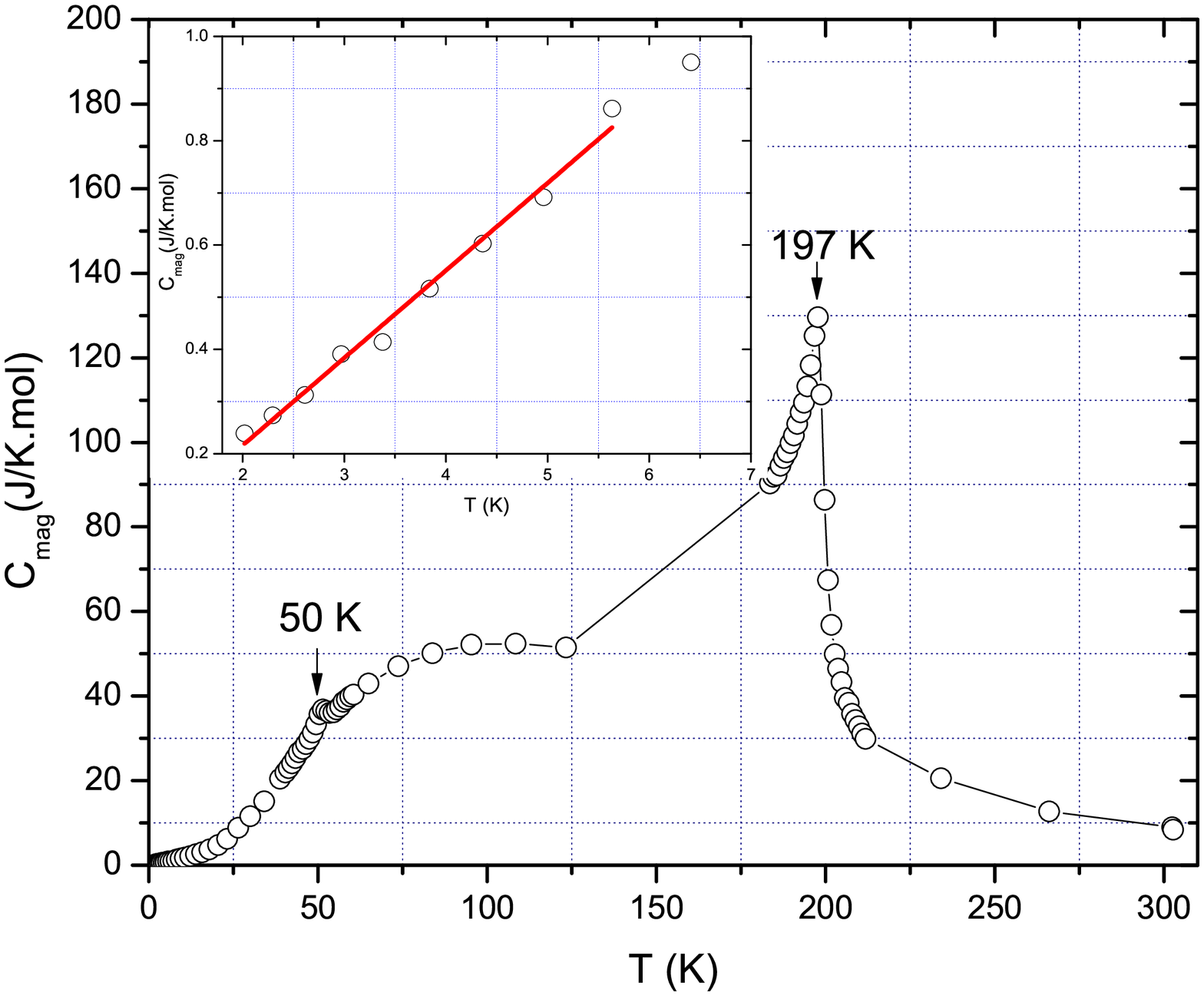}%
}
\subfloat[\label{sfig:Magnetic_entropy_esimation}]{%
  \includegraphics[scale=0.35]{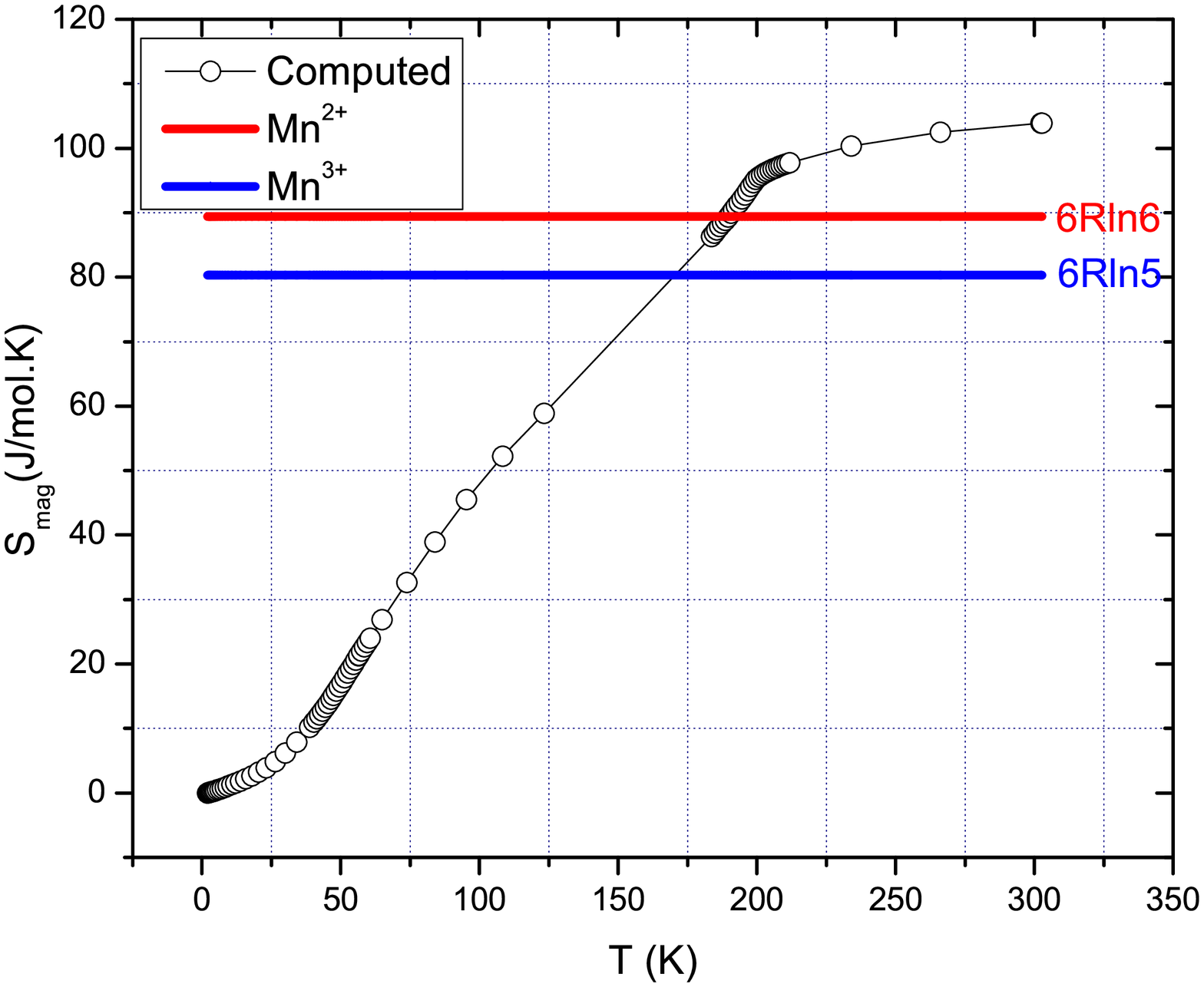}%
}
\caption{a) Zero magnetic field heat capacity of Mn$_6$Ni$_{16}$Si$_7$ (open circles) and the lattice match compound Ti$_6$Ni$_{16}$Si$_7$ (closed circles) from 2 K to 303 K. Mn$_6$Ni$_{16}$Si$_7$ shows a weak transition at 50 K, which is magnified in the inset and a sharp antiferromagnetic to paramagnetic transition at 198 K. Ti$_6$Ni$_{16}$Si$_7$ also undergoes a transition at 142 K. $ b)$Magnetic contribution, $C_{mag}$ for Mn$_6$Ni$_{16}$Si$_7$ as function of temperature. The magnetic contribution, $C_{mag}$ is obtained by direct subtraction of the lattice match compound Ti$_6$Ni$_{16}$Si$_7$ data ignoring the phase transition. A linear fitting of $C_{mag}$ below 6 K (magnified in the inset) supports the spin freezing anomaly.  c) Entropy, $S_{mag}$ over the whole temperature range (2 K-303 K) is compared with $6Rln6$ for $S=\frac{5}{2}$ Mn$^{2+}$ and $6Rln5$ for $S=2$ Mn$^{3+}$.}
\label{fig:HeatCapacity}
\end{figure}

Fig.~\ref{sfig:Heat_Capacity_Comparison} shows the heat capacity of Mn$_6$Ni$_{16}$Si$_7$ measured under zero magnetic field between 2 K and 303 K. The data clearly depict a $\lambda$ anomaly peak at 197 K that corresponds to a transition between antiferromagnetic and paramagnetic states. A second transition at 50 K characterized by a weak, almost undetectable peak in heat capacity data (inset in Fig. \ref{sfig:Heat_Capacity_Comparison}), is consistent with transition observed in the magnetic susceptibility plots in Fig.~\ref{fig:ZFC-FC-100-110-111-1000Oe}. To obtain the magnetic contribution to the heat capacity of Mn$_6$Ni$_{16}$Si$_7$ one needs to subtract that of a suitable lattice match, normally a paramagnetic iso-structural compound~\cite{Marjerrison2016}. A previous study of the magnetic properties of Ti$_6$Ni$_{16}$Si$_7$  showed a Pauli paramagnetic like behaviour \cite{Holman2008}, which was also observed in our measurements of magnetic susceptibility of the compound under an applied magnetic field of 1000 Oe (SI-3). However, the heat capacity as a function of temperature showed a cusp at 140 K for Ti$_6$Ni$_{16}$Si$_7$  (closed circles in Fig. \ref{sfig:Heat_Capacity_Comparison}), which is an indication of a phase transition, nevertheless unexpected, given the data of \citet{Holman2008} and our magnetic measurements. The origin of this anomaly is unclear at present, but apart from this feature, Ti$_6$Ni$_{16}$Si$_7$ appeared to be a reasonable lattice match. Consequently, the magnetic contribution of the heat capacity for Mn$_6$Ni$_{16}$Si$_7$ was obtained (Fig.~\ref{sfig:CmagT}) by a direct subtraction of the Ti$_6$Ni$_{16}$Si$_7$ data. Note that the 50 K anomaly is now clearly evident. Furthermore, a linear fitting of the C$_{mag}$ was obtained below 6 K (inset in Fig.~\ref{sfig:CmagT}. Such a linear dependence of heat capacity is usually a sign of spin freezing~\cite{Ramirez1994} which further supports the cusp at 6 K to be associated with such an anomaly. To compute the expected entropy loss for this material requires selection of an oxidation state for Mn. For Mn$^{3+}$ S$_{mag}$ = 80.28 J/mol-K while for Mn$^{2+}$ it is 89.38 J/mol.K. The computed magnetic entropy (Fig.~\ref{sfig:Magnetic_entropy_esimation}) however, showed S$_{mag}$ of 17.14 J/mol.K at 50 K, 94.03 J/mol.K at 197 K and, 103.9 J/mol.K at 303
 K, which exceeds reasonable values for this material, assuming that Mn is the only magnetic atom. The data of \citet{Holman2008} suggests that there is no ordered moment on the Ni sites in this class of materials. The discrepancy might then be attributed to the choice of Ti$_6$Ni$_{16}$Si$_7$ as a lattice match which apparently underestimates the lattice contribution in Mn$_6$Ni$_{16}$Si$_7$.

\subsection{Neutron diffraction}\label{sec:Neutron diffraction}

\begin{figure*}
\subfloat[\label{fig:Neutron_298K_match}]{%
   \includegraphics[scale=0.35]{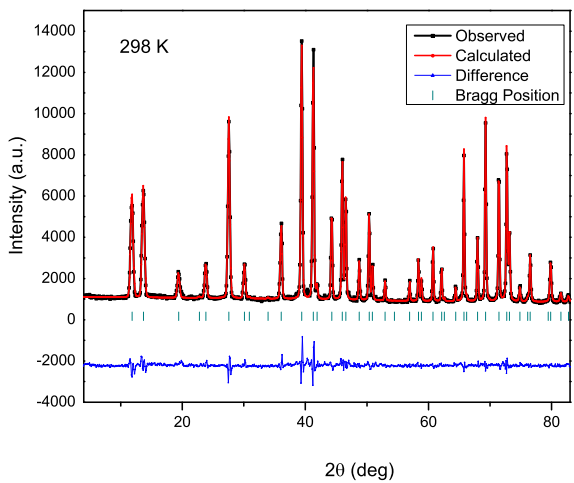}%
}
\subfloat[\label{fig:Neutron_100K_match}]{%
    \includegraphics[scale=0.35]{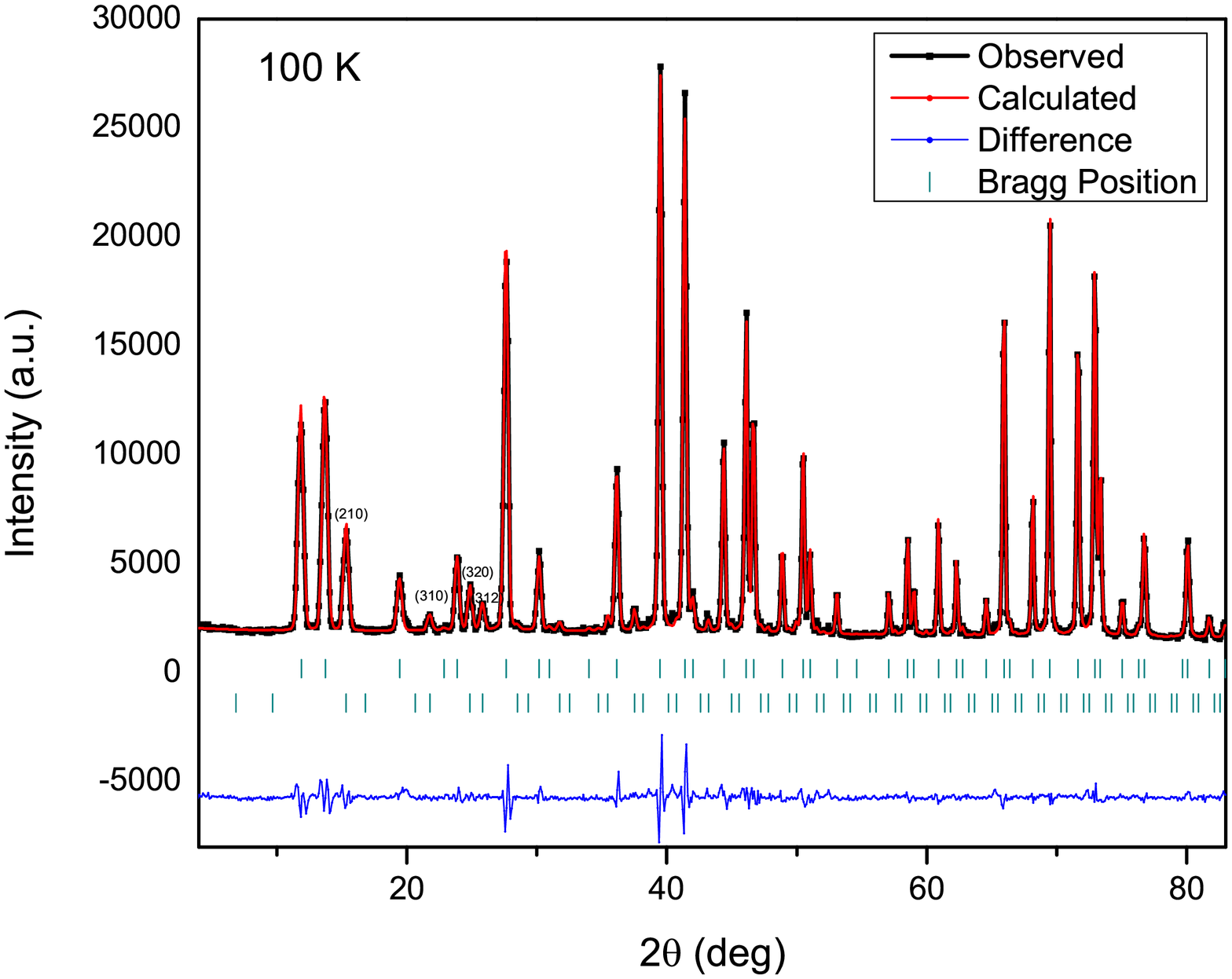}%
 }\hfill
\subfloat[\label{fig:Neutron_4K_match}]{%
 \includegraphics[scale=0.35]{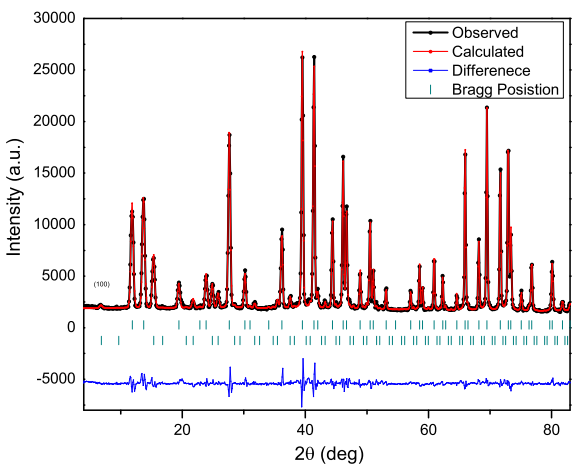}
 }
 \subfloat[\label{fig:Neutron_100_peak}]{%
  \includegraphics[scale=0.35]{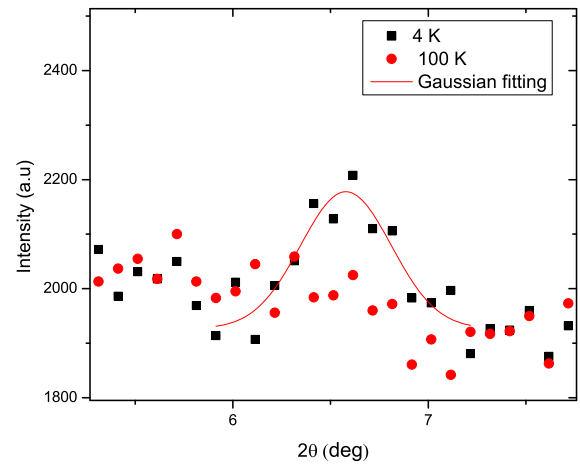}
 }
 \caption{Rietveld refinement of the neutron powder diffraction data at a) 298 K, b) 100 K and c) 4 K. (d) A comparison of 4 K and 100 K neutron data at low $2\theta$ showing (100) peak at 4 K.}
 \label{fig:NeutronDiffractionmatch}
\end{figure*}

\begin{figure}
\centering
   \includegraphics[scale=1.7]{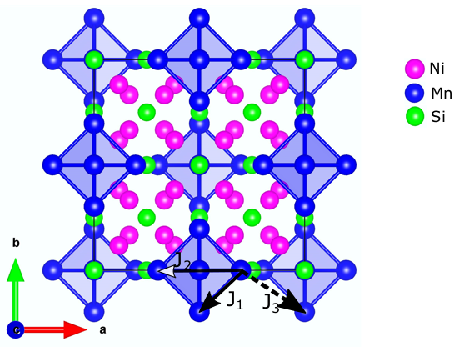}\\
    \caption{Refined crystal structure in the paramagnetic state. In the structure, Mn ions are connected by 4$J_1$, 1$J_2$ and 4$J_3$ bonds.}
      \label{fig:298K_crystal}
\end{figure}

Neutron diffraction data were obtained on a polycrystalline powder sample at 298 K, 100 K and 4 K corresponding to the three different regimes of magnetic susceptibility and heat capacity behaviour. As indicated by magnetic measurement, at room temperature the structure was refined to be a paramagnetic type with no additional magnetic contribution (Fig. \ref{fig:Neutron_298K_match}). The room temperature structure was refined by the Rietveld method (Fig. \ref{fig:Neutron_298K_match}), and the structural parameters are listed in Table~\ref{Table:Ni16Mn6Si7_refined}. The structural parameters were found to be in excellent agreement with previously reported results for Mn$_6$Ni$_{16}$Si$_7$ \cite{STADNIK1983,Kolenda1991,Yan2009}. The occupancies of individual elements were kept constant during the refinement. The Mn atoms occupy the sites of regular octahedra which are not directly connected with other octahedra. Each Mn atom within an octahedron  is coupled with four nearest neighbours by exchange constant $J_1$, one next nearest neighbour by $J_2$ and with four neighbours of the adjacent octahedra by $J_3$, as shown in figure \ref{fig:298K_crystal}. The $J_1$, $J_2$, and $J_3$ pathway lengths are approximately 0.279 $\mathrm{a}$, 0.393 $\mathrm{a}$ and 0.429 $\mathrm{a}$, respectively, where $\mathrm{a}$ is the lattice constant.\\

\begin{table}[htp]
    \caption{Refined structural parameters for Mn$_6$Ni$_{16}$Si$_7$}\label{Table:Ni16Mn6Si7_refined}
     \centering
        \begin{tabular}{c c c c c}
        \hline
        \hline
        &Temperature& 298 K &100 K&4 K\\
        \hline
        &Lattice constant(${\AA}$)&11.1497(4)&11.1238(3)&11.1198(4)\\
        &Spacegroup&Fm$\bar{3}$m&Fm$\bar{3}$m&Fm$\bar{3}$m\\
        &Magnetic phase&Paramagnetic&Antiferromagnetic&Canted Antiferromagnetic\\
        &Mn; 24e (x, 1/2, 1/2)&0.6969(3)&0.6969(3)&0.6969(3)\\
        &B(Mn) (${\AA}^2$)&0.49(15)&0.32(14)&0.09(5)\\
        &M(Mn-I)($\mu_B$)&-&4.45(5)&4.74(11)\\
        &M(Mn-II)($\mu_B$)&-&0&0.65(24)\\
        &Ni1; 32f (x, x, x)&0.33322(11)&0.33322(11)&0.33322(11)\\
        &B(Ni1) (${\AA}^2$)&0.58(5)&0.36(5)&0.22(5)\\
        &M(Ni1)($\mu_B$)&-&0&0\\
        &Ni2; 32f (x, x, x)&0.11780(12)&0.11780(12)&0.11780(12)\\
        &B(Ni2) (${\AA}^2$)&0.58(5)&0.49(5)&0.36(5)\\
        &M(Ni2)($\mu_B$)&-&0&0\\
        &Si1; 4a (1/2, 1/2, 0)&-&-&-\\
        &B(Si1) (${\AA}^2$)&0.48(10)&0.74(27)&0.66(28)\\
        &Si2; 24e (1/2, 1/4, 1/4)&-&-&-\\
        &B(Si2) (${\AA}^2$)&0.48(10)&0.35(10)&0.34(10)\\
        &$\chi^2$&5.33&9.66&10.4\\
        &$R_{wp}$&11.4&10.3&10.7\\
        &$R_{F}$&2.51&1.66&1.94\\
        &$R_{mag}$&-&8.81&8.06\\
                \hline
        \end{tabular}
\end{table}
\begin{figure}
\centering
   \includegraphics[scale=0.7]{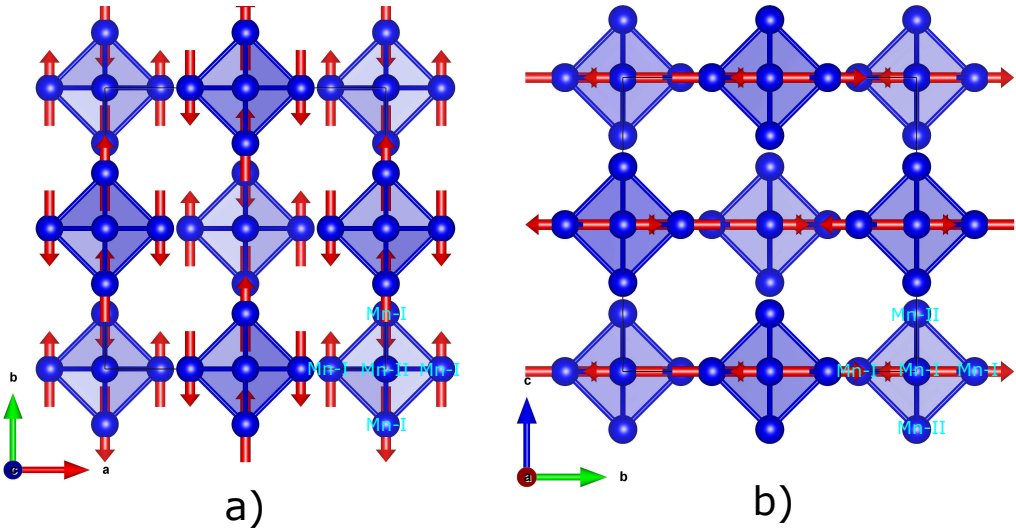}\\
    \caption{ Magnetic structure at 100 K showing a two-dimensional (2D) magnetic arrangement magnetic moments between Mn-I ions (parallel to $\mathrm{ab}$ plane in the current coordinate system). Mn-II ions remain frustrated. The structures shown in the figure represent projection along the $\mathrm{ab}$ and b) $\mathrm{bc}$ planes, respectively.}
      \label{fig:100k_magnetic_structure}
\end{figure}

The neutron diffraction data collected at 100 K show new reflections, relative to room temperature, which can be indexed as (210), (310), (320) and (312). These reflections are systematically absent in $Fm-3m$ symmetry structure and can thus be associated with magnetic ordering. The same reflections were reported by \citet{Kolenda1991} in their studies of the compound.\\

The solution of the magnetic structure was based on the representation analysis method using SARAh~\cite{Wills2000}. For the ordering, wave vector  k = (0 0 1) four  basis vectors $\Gamma_2$, $\Gamma_3$, $\Gamma_9$ and $\Gamma_{10}$ resulted from this approach. The best refinement was obtained using $\Gamma_9$ (refinement parameters for all $\Gamma$ configurations are listed in SI-table II). The Rietveld refinement profile is shown in Fig.~\ref{fig:Neutron_100K_match}, and magnetic moments are listed in table \ref{Table:Ni16Mn6Si7_refined}. The resulting magnetic structure (Fig.~\ref{fig:100k_magnetic_structure}) shows a 2D antiferromagnetic configuration of Mn$^{2+}$ atoms involving only four of the six sites on the Mn$_{6}$ octahedron oriented along $<100>$ directions in an ab plane. Here, Mn occupancy is separated into 2 Mn sublattices, Mn-I and Mn-II. Mn-I contain chains of atoms on a principal plane arranged antiferromagnetically with the nearest neighbour atoms. Mn-II atoms lie on top and bottom of the Mn-I chains and remain paramagnetic providing evidence of spin frustration within each octahedron. It can be seen that two of four $J_1$ Mn are coupled antiferromagnetically while the remaining two are paramagnetic. Mn atoms connected by $J_2$ are coupled ferromagnetically as would be required given dominant nearest neighbour antiferromagnetic exchange. It can also be noticed that two-dimensional planes formed by the ordered Mn ions are coupled antiferromagnetically with each other indicating strong antiferromagnetism in Mn$_6$Ni$_{16}$Si$_7$. The Mn-I moments were refined to be 4.45 (5) $\mu_B$ per Mn ion. It can be seen that the moment is close to the theoretical and experimentally reported value for Mn$^{2+}$, i.e., 5 $\mu_B$ (with g=2 and $S=\frac{5}{2}$) and 4.7 $\mu_B$ \cite{Fries1997}, respectively. The moment for the Nickel ions were refined to be 0 to within 2$\sigma$. It can be seen that our model is different from the one proposed by~\citet{Kolenda1991}. Refinement of the 100 K data using their proposed magnetic structure showed relatively poor agreement with the observed powder pattern (SI-table II and SI-4). On the other hand, our model of Fig.~\ref{fig:100k_magnetic_structure}, with frustrated Mn spins is consistent with the observed bulk susceptibility data which shows a paramagnetic like contribution persisting below $T_{N}$ = 198 K.\\

\begin{figure}
\centering
   \includegraphics[scale=0.7]{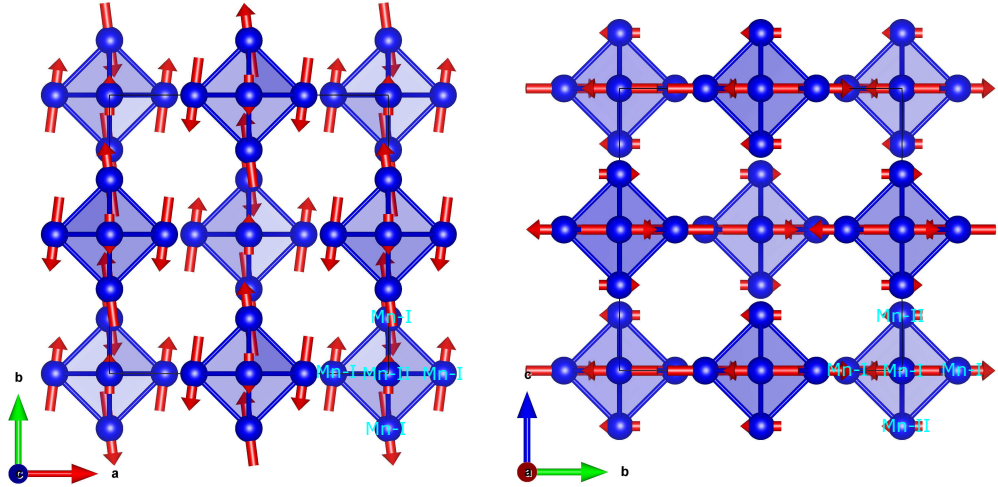}\\
    \caption{A hidden spin canted antiferromagnetic structure is favored at 4K. The two dimensional magnetic symmetry is broken and the moments are canted in ($\mathrm{ab}$) plane. Magnetic moment vectors are guide to eye only. The structures presented in the figure corresponds to projection along the $\mathrm{ab}$ and b) $\mathrm{bc}$ planes, respectively.}
      \label{fig:4k_magnetic_structure}
\end{figure}
A careful comparison of neutron diffraction at 4 K and 100 K revealed an extra (100) peak at 4 K (Fig.~\ref{fig:Neutron_100_peak}). The presence of such a reflection implies the magnetic moments cannot be parallel to (100). Consequently, it can be presumed that the 2D antiferromagnetic symmetry is broken. This result can be associated with the phase transition at 50 K. The refinement showed that the Mn-I moment value at 4 K is increased slightly to 4.74 (11) $\mu_B$  and that the spins make a canting angle with respect to (100) of 9 (1$^\circ$) (Fig.~\ref{fig:4k_magnetic_structure}). This is sometimes called hidden spin canting which refers to a configuration where the moments of multiple sublattices are canted about the axis but still in a collinear arrangement~\cite{Engelfriet1980,Carlin1981,Tian2003,Wang2005}. Note that the moments are still consistent with values for the $S=\frac{5}{2}$ system \cite{Fries1997}. As well, a small moment of 0.65 (24) $\mu_B$ appears on the Mn-II site which may be real. Note that even below 50 K, the model predicts a high concentration of paramagnetic spins, consistent with the bulk susceptibility which continues to increase at lower temperatures. The relevant refinement parameters are shown in Table \ref{Table:Ni16Mn6Si7_refined} and the Rietveld profile is shown in Fig.~\ref{fig:Neutron_4K_match}.

\section{Discussion}\label{sec:Discussion}

The observation of the magnetic susceptibility and heat capacity data suggesting a paramagnetic system above 197 K is reflected well in the neutron refinement at 298 K where no magnetic ordering of the Mn$_6$Ni$_{16}$Si$_7$ system was observed. The magnetic structure obtained in the 100 K refinement found Mn-II ions to be remaining as paramagnetic, while Mn-I atoms are antiferromagnetically coupled suggesting the system be a 2D geometrically frustrated antiferromagnet. The configuration can be related to the magnetic susceptibility data from 197 K to 50 K. Although the negative $\Theta_{CW}$ from the Curie-Weiss fitting of the inverse susceptibility confirmed the system to be antiferromagnetic below 197 K. Also, the magnetic susceptibility data showed to be consisting a paramagnetic contribution that can be explained in terms of geometrically frustrated  Mn-II atoms. The refinement at 4 K showed the breaking of the two dimensional symmetry yielding a hidden spin-canted magnetic configuration. The structure can be associated with the phase transition at 50 K detected by the magnetic susceptibility and heat capacity measurement. Note that, the system also demonstrates another phase anomaly below 6 K, due to the spin freezing effect. Although the refinement at 4 K found the Mn-II to be ordering with small moments, the system still contains higher concentrations of paramagnetic Mn ions. Freezing of these weak moments that can result in the rapid drop of ZFC magnetization below 6 K, is quite possible. Such a weak phenomenon however, can not be detected by the neutron diffraction experiments.\\

\section{Conclusions}\label{sec:Conclusion}

In this work, it has been shown that Mn$_6$Ni$_{16}$Si$_7$ shows remarkably complex magnetic behaviour. Three phase transitions occur upon cooling from ambient temperature. Below 197 K, long range anti-ferromagnetic order occurs but with only 2/3 of the Mn spins involved, which can be understood in terms of geometric magnetic frustration within the Mn$_{6}$ octahedra. The Mn moments are oriented along $<100>$ directions in a planar pattern. Below 50 K, these ordered moments cant away from $<100>$ by an angle of 9$^\circ$ but remain collinear. Simultaneously, the previously paramagnetic Mn sites develop a small moment of 0.6 $\mu_B$, compared with 4.70 $\mu_B$ found on the other Mn atoms. Finally, below 6 K, there is evidence of spin freezing involving the remaining paramagnetic Mn spins.

\begin{acknowledgement}
Authors would like to thank Dr. A. S. Wills (University College London) for stimulating discussions. Dr. A. Dabkowski (McMaster University) assisted with the crystal growth process. Financial support of Natural Sciences and Engineering Research Council of Canada under the NSERC grant: "Artificially Structured Multiferroic Composites based on the Heusler alloys" is gratefully acknowledged.
\end{acknowledgement}

\providecommand{\latin}[1]{#1}
\makeatletter
\providecommand{\doi}
  {\begingroup\let\do\@makeother\dospecials
  \catcode`\{=1 \catcode`\}=2 \doi@aux}
\providecommand{\doi@aux}[1]{\endgroup\texttt{#1}}
\makeatother
\providecommand*\mcitethebibliography{\thebibliography}
\csname @ifundefined\endcsname{endmcitethebibliography}
  {\let\endmcitethebibliography\endthebibliography}{}

\begin{suppinfo}

\begin{figure}[htp]
\centering
   \includegraphics[scale=0.5]{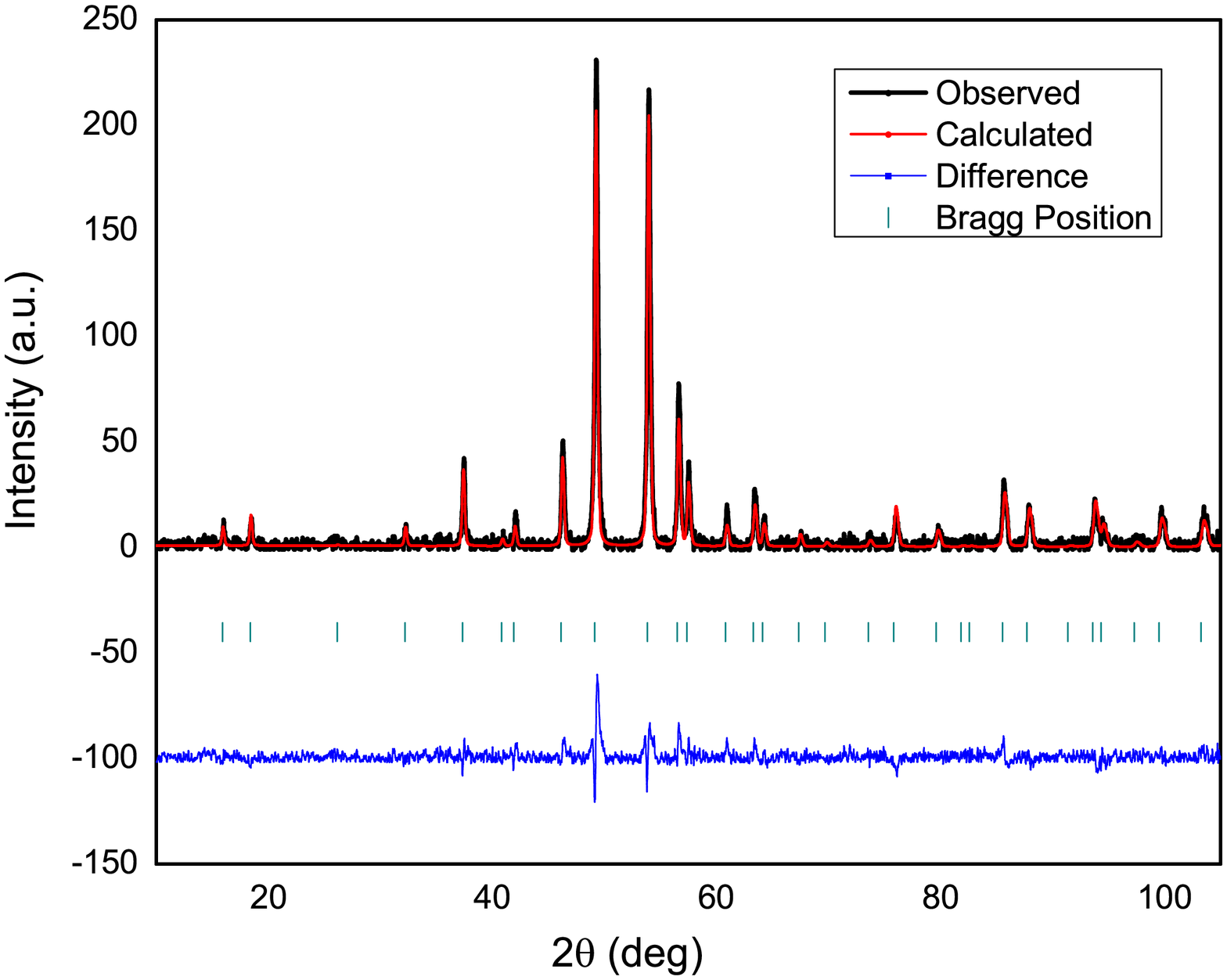}\\
   \caption{Rietveld refinement of the XRD powder diffraction data at 298 K  Ni$_{16}$Mn$_{6}$Si$_{7}$}\label{fig:XRD_298K_match}
\end{figure}

\begin{figure}[!ht]
\centering
   \includegraphics[scale=0.5]{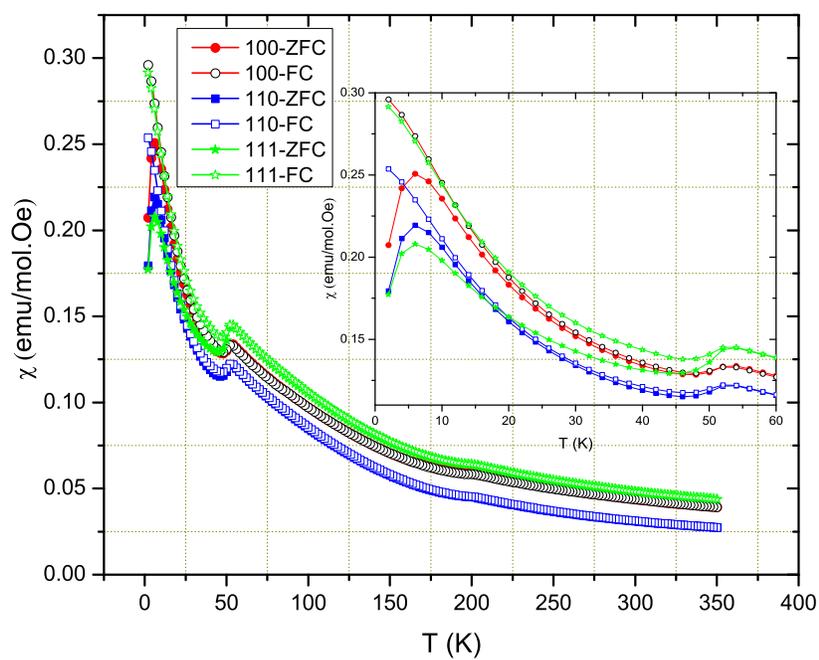}\\
   \caption{ZFC-FC magnetic susceptibility as a function of temperature with a magnetic field of 1000 Oe is applied in $<100>$, $<110>$ and $<111>$ directions. Inset shows anisotropic ZFC-FC magnetic susceptibly behavior below the 50 K transition.}
   \label{fig:ZFC-1000Oe-all}
\end{figure}

\begin{table}[htp]
    \caption{Anisotropy of the FC Curie-Weiss parameters.}\label{Table:Ni16Mn6Si7_C-W}
     \centering
        \begin{tabular}{c c c c c}
        \hline
        \hline
        &Field direction &$C$ (emu.K/mol.Oe) &$\mu_{eff}$/Mn ($\mu_B$)&$\Theta_{CW} (K)$ \\
        \hline
        &$<100>$&17.75 (4)&4.83 (24)&-105 (1)\\
        &$<110>$&10.20 (4)&3.65 (22)& -26 (1)\\
        &$<111>$&20.38 (10)&5.17 (36) &-120 (2)\\

                \hline
        \end{tabular}
\end{table}

\begin{figure}[htp]
\centering
   \includegraphics[scale=0.5]{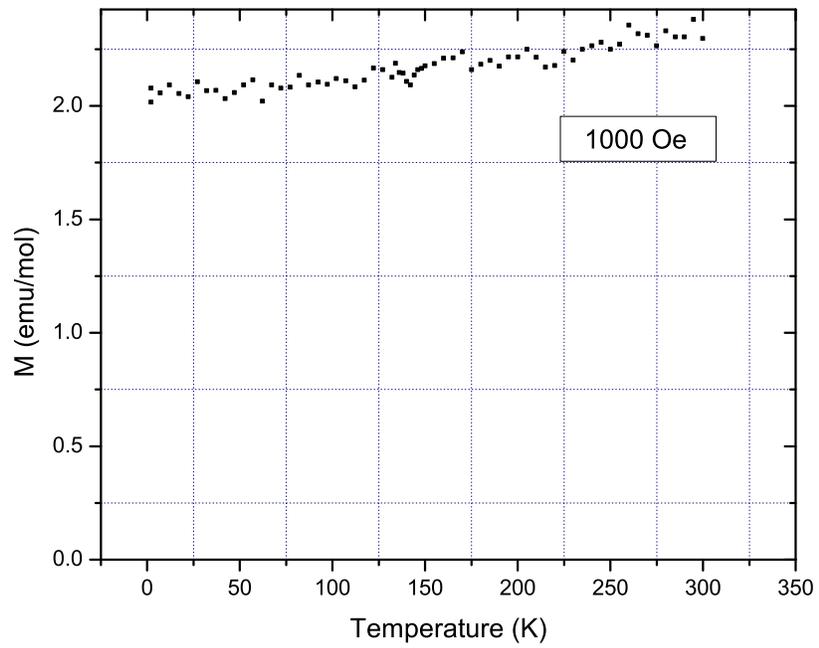}\\
   \caption{ZFC magnetic susceptibility of Ni$_{16}$Ti$_{6}$Si$_{7}$ under a constant magnetic field of 1000 Oe.}\label{fig:Ni16Ti6Si7-1000Oe-ZFC}
\end{figure}

\begin{table}[htp]
    \caption{Magnetic refinement at 100 K for different $\Gamma$ configurations}\label{Table:Ni16Mn6Si7_Gamma}
     \centering
        \begin{tabular}{c c c c c}
        \hline
        \hline
        &Configurations& $R_{mag}$ &$\chi^2$ &$R_{wp}$\\
        \hline
        &$\Gamma_{2}$&21.18&11.4&11.4\\
        &$\Gamma_{3}$&52.24&20.6&15.3\\
        &$\Gamma_{9}$&8.81&9.66&10.3\\
        &$\Gamma_{10}$&97.08&35.6&20.2\\
        &Previously published~\cite{Kolenda1991}&62.95&23.9&16.5\\

                \hline
        \end{tabular}
\end{table}

\begin{figure}[htp]
   \includegraphics[scale=0.5]{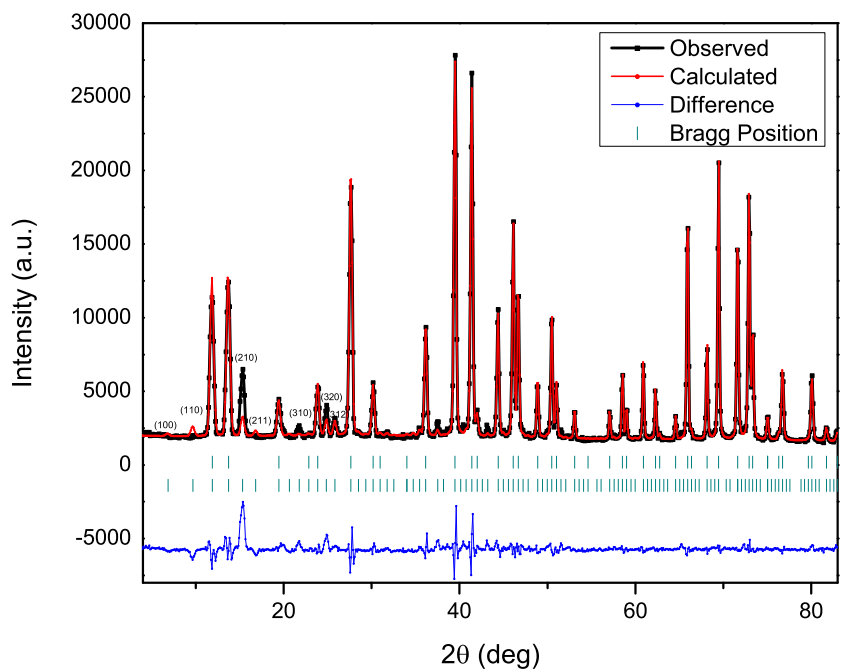}\\
   \caption{Refinement of neutron diffraction pattern with the magnetic structure model proposed by \citet{Kolenda1991}}\label{fig:Koledna_simulation}
\end{figure}
\end{suppinfo}

\end{document}